\documentstyle[epsf,12pt]{article}
\title{\hspace{10cm} {\scriptsize DPNU-99-15}\\
Lie-Group Approach to Perturbative Renormalization-Group Method}

\author{Shin-itiro Goto ,Yuji Masutomi and Kazuhiro Nozaki \\
        Department of Physics, Nagoya University\\
          Nagoya 464-8602, Japan\\}

\makeindex

\begin{document}

\baselineskip 2.5ex
\parskip 3.5ex
\maketitle
\renewcommand{\thesection}{\arabic{section}}
\renewcommand{\thesubsection}{\arabic{section}.\Alph{subsection}}
\renewcommand{\theequation}{\arabic{section}.\arabic{equation}}
\newcommand{\eps}{\epsilon}
\newcommand{\A}{\tilde A}
\newcommand{\B}{\tilde B}
\newcommand{\p}{\tilde p}
\newcommand{\h}{\tilde h}
\newcommand{\pvt}{\tilde{\bm p}}
\newcommand{\pv}{\bm p}
\newcommand{\rv}{\bm r}
\newcommand{\etav}{\bm {\eta}}
\newcommand{\phit}{\tilde{\phi}}
\newcommand{\rvp}{{\bm r}_{\perp}}
\newcommand{\ad}{\dot a}
\newcommand{\add}{\ddot a}
\newcommand{\bm}[1]{\mbox{\boldmath $#1$}}
\newcommand{\svec}[1]{\mbox{{\footnotesize $\bm{#1}$}}}
\renewcommand{\vec}{\bm}
\newcommand{\der}[2]{\frac{d {#1}}{d {#2}}}
\newcommand{\beq}{\begin{equation}}
\newcommand{\beqa}{\begin{eqnarray}}
\newcommand{\eeq}{\end{equation}}
\newcommand{\eeqa}{\end{eqnarray}}
\newcommand{\la}{\langle}
\newcommand{\ra}{\rangle}
\renewcommand{\vec}{\bm}
\newcommand{\un}{\underline}
\newcommand{\noi}{\noindent}
\newcommand{\lb}{\label}
\newcommand{\fr}[1]{(\ref{#1})}
\newcommand{\non}{\nonumber}
\newcommand{\fns}{\footnotesize}
\newcommand{\map}{\rightarrow}
\newcommand{\maps}{\rightarrow}
\newcommand{\imply}{\Rightarrow}
\newcommand{\implies}{\Rightarrow}

\begin{abstract}
The Lie-group approach to the perturbative renormalization group (RG)
method is developed to obtain an asymptotic solutions of both autonomous
 and non-autonomous ordinary differential equations. Reduction of some
 partial differetial equations to typical RG equations is also
achieved by this approach and a simple recipe for providing RG equations
is presented.
\end{abstract}

\pagebreak

\section{Introduction}

A novel method based on the perturbative renormalization group
 (RG) theory has been developed as an asymptotic singular
perturbation technique \cite{cgo0} and has been succesively applied to
ordinary differential equations (ODE)  \cite{cgo1} \cite{yama}
and some partial differential equations (PDE)
\cite{cgo1}\cite{sasa}\cite{mano1}. An approach from a geometrical point of
 view was also proposed, which is called an envelope method \cite{kuni}.\\
The renormalization group method removes secular or divergent
terms from a naive perturbation series by renormalizing integral constants
 i.e. a renormalization transformation and reaches a slow-motion and/or
 large-scale equation as a renormalization group(RG) equation.
In this reduction of an asymptotic equation, it is not
necessary to introduce slow or large scales a priori in contrast to
the other methods such as the reductive perturbation method \cite{tani}
and various multiple-scale methods \cite{nay}. Instead, a
renormalization transformation determines an asymptotic equation as a
RG equation.
Here, it is crucial to extract the structure of secular terms
from a perturbation series in order to obtain a renormalization
transformation.
For ODE, the structure of secular terms are  often relatively
simple and, as shown in this paper,  it is possible to calculate explicit
secular solutions up to arbitrary order in principle and obtain
 an asymptotic form of the renormalization transformation of all order
 in some examples .
However, for PDE, there are many divergent solutions in a perturbed equation
and it is ambiguous which divergent solutions should be renormalized away
\cite{mano2}. Although an attempt to overcome this difficulty was
presented \cite{mano1}, it is far from satisfactory. \\
The group structure of the renormalization transformation has
not yet been necessarily clear in the previous applications of
the RG method and it is desireble to make the renormalization procedure
more transparent for its extended applications.\\
In this paper, we construct a representation of the Lie group from a
 renormalization transformation in several examples and identify the
 procedure of renormalization with one to derive an asymptotic
 expression to a generator of the Lie group.
 Although the view of the Lie group has been introduced
 to the RG method in  systems with a translational symmetry with respect to
 a independent variable \cite{shir}\cite{sasa},
 there are few examples based on the Lie group theory and only
 the leading or low order RG equations are derived \cite{mano1} .
 For the purpose of obtaining higher-order RG equations and applying
 the RG method to wider systems such as non-autonomous systems without a
 translational symmetry and PDE, much more investigations are still
 necessary to refine the Lie-group approach. \\
 In Section {\bf 2}, we present some clear examples, in which  an
asymptotic expression to a generator of the Lie group
(a renormalization transformation)
  is obtained up to arbitrary order, in principle, by naive pertubation
  calculations. In the simplest example in Section {\bf 2.A}
  , an asymptotic expression of a generator is shown to converge and
  we recover an exact solution. In Section {\bf 2.B} ,
  we extend the Lie-group approach
  to a non-autonomous system of ODE by introducing a simple shift
  operation. Through these simple examples, we present a recipe for
  obtaining an asymptotic representation to the Lie group underlying
  the original system. In the remaining parts of Section {\bf 2},
   we give some physically interesting examples. In an example in
   Section {\bf 2.C}, a renormalization transformation has a translational
    symmetry though an original ODE does not have one.
  In Section {\bf 2.D}, we present a peculiar example in the Einstain's
  gravitational equation where an original ODE and its leading
  order RG equation are autonomous but a higher order RG equation
  becomes non-autonomous. Even in this peculiar case, our Lie-group
  approach works well and
  gives an asymptotic solution describing an expanding space.
  In Section {\bf 2.E}, it is shown that secular solutions to
   perturbed equations for a weakly nonlinear oscillator can be removed
   asymptotically up to all order. We  also derive  adiabatic equations
   for parameters of the K-dV soliton under perturbations
   in Section {\bf 2.F}.\\
 In Section {\bf 3}, an extension of the RG method in terms of the
  Lie group is presented for some nonlinear PDE. We derive two typical
  soliton equations such as the nonlinear Schr\"odinger equation and the
  Kadomtsev-Pitviashvili equation with higher order corrections as RG
  equations. As the final example,
  phase equations for interface and oscillating solutions are derived from
  a general reaction-diffusion system.\\
  In Section {\bf 4}, we give a simple recipe for providing RG equations as
  a summary of this paper.

\setcounter{equation}{0}
\section{ Ordinary Differetial Equations}

\subsection{Linear Autonomous System}

As an example in which a generator to the Lie group of a renormalization
transformation is constructed exactly, we discuss a boundary layer type
problem \cite{nay}.
\beq
\left(\epsilon \der{^2}{x^2} +\der{}{x} + 1\right) u =0, \lb{ble}
\eeq
where $u(0)=0$. Since a small parameter $\eps$ multiplying the highest
derivative causes a  severe singularity at a boundary $x=0$, we introduce
a transformation $x=\eps t$. Then, \fr{ble} reads
\beq
\left( \der{^2}{t^2} +\der{}{t} + \eps \right) u =0, \lb{ble2}
\eeq
which is called the inner equation in the boundary layer problem.
Expanding $u$  as
\beq
u =A +Be^{-t} +\epsilon u_1+\epsilon^2 u_2+\cdots,\non
\eeq
we have
\beq
L(t)u_n\equiv \left(\der{^2}{t^2} +\der{}{t}\right)u_n = -u_{n-1},\non
\eeq
where $u_0=A +Be^{-t}$ and $A, B$ are arbitrary constants.
Secular solutions can be written as
\beq
u_n = AP_n(t)+BP_n(-t)e^{-t},\non
\eeq
where $P_0=1$ while  $P_n(t) \quad (n\ge 1)$ is a polynomial solution
of
\beqa
L(t)P_n&=&-P_{n-1}, \non\\
P_1&=&-t,\quad P_2=t^2/2-t,\quad P_3=-t^3/6+t^2-2t.\non
\eeqa
Here, we observe that  a secular term $P_n(t)$ is a polynomial of degree
$n$. In order to eliminate these secular terms by renormalizing integral
constants $A, B$, the following naive renormalization transformations
 $A \to \A(t)$ and $B \to \B(t)$ are introduced as
\beqa
\A(t)&=&A(1+\eps P_1(t)+\eps^2 P_2(t)+\cdots),\lb{rentr1}\\
\B(t)&=&B(1+\eps P_1(-t)+\eps^2 P_2(-t)+\cdots).\lb{rentr2}
\eeqa
Note that this definition of renormalization transformations is
different from one in \cite{cgo1}.
 In order to obtain  a representation to
the Lie group from \fr{rentr1}, we take advantage of a translational
symmetry in \fr{ble2}.  It is easy to see that $\A(t), \B(t) $ also
 enjoy a translational symmetry ,that is
\beq
 L(t)\A+\eps\A=0, \quad L(-t)\B+\eps\B=0. \lb{auab}
\eeq
The renormalization transformation \fr{rentr1} is interpreted as a
Taylor series for the solution $\A$ of \fr{auab} around $t=0$,
while the origin of expansion can be
shifted to arbitrary positions by virtue of a translational symmetry.
Thus, we obtain an asymptotic representation to the Lie group $G_\tau$
immediately from \fr{rentr1}.
\beqa
\A(t+\tau)&=&\A(t)(1+\eps P_1(\tau)+\eps^2 P_2(\tau)+\cdots),\non\\
&\equiv&G_\tau \A(t).\lb{autolie}
\eeqa
In terms of a generator $\tau \partial_t$ of the Lie group,
$\A(t+\tau)$ is also expanded as
\beqa
\A(t+\tau)&=&\exp(\tau\partial_t)\A(t),\non \\
&=&(1+ \tau\partial_t+({\tau^2}/2){\partial_t}^2+\cdots)\A(t).\lb{aulig}
\eeqa
Equating coefficients of equal powers of $\tau$ in \fr{autolie} and
\fr{aulig}, we have an asymptotic form of a generator  of the Lie group
called a RG equation
\beqa
\der{\A}{t} &=&\partial_\tau[G_\tau \A(t)]_{\tau=0},\non \\
&=&\partial_\tau[\eps P_1(\tau)+\eps^2 P_2(\tau)
+\cdots]_{\tau=0}\A,\non \\
&=&-\eps(1+\eps+2\eps^2+\cdots)\A\equiv \eps\delta\A,\lb{augen}
\eeqa
All other equations for higher order derivatives are
guranteed to be consistent with \fr{augen} by virtue of the theory of
Lie group or arbitrainess of $\tau$.
Since $\delta=-(1+\eps+2\eps^2+\cdots)$ converges to
$(-1+\sqrt{1-4\eps})/(2\eps)$ for $\eps<1/4$, \fr{augen} gives an exact
 form of the generator. Applying a similar analysis to \fr{rentr2},
 we obtain  a RG equation for $\B$:
\beqa
\der{\B}{t} &=&\partial_\tau[\eps P_1(-\tau)+\eps^2 P_2(-\tau)
+\cdots]_{\tau=0}\B,\non \\
&=&\eps(1+\eps+2\eps^2+\cdots)\B=-\eps\delta\B.\non
\eeqa
Thus,renormalizing all secular terms in the perturbation series,
 we recover an exact solution of  \fr{ble} with the boundary
condition $u(0)=0$.
\beq
u=\A(0)\exp(\delta x)+\B(0)\exp(-\frac{x}{\eps}-\delta x),\non
\eeq
where $\A(0)+\B(0)=0$.

\subsection{Oscillator with time-dependent spring constant}

To extend the Lie-group approach described in the previous section
to systems without a translational symmetry with respect to $t$, we
consider an linear but nontrivial oscillator governed by
\beq
\der{^2 u}{t^2} + u = \epsilon t u,\non
\eeq
which was analyzed in \cite{cgo1} to the first order in $\eps$.
Here, we try to extend the RG analysis to arbitrary order in terms of the
Lie group.
Expanding $u$ as
\beq
u =A e^{-it} +\epsilon u_1+\epsilon^2 u_2+\cdots+\mbox{c.c.},\non
\eeq
we have
\beq
\left(\der{^2 }{t^2} + 1\right)u_n = tu_{n-1},\non
\eeq
where c.c. stands for the complex conjugate of the preceding expression,
$n=1,2,\cdots$ and $u_0=A e^{-it}$.
Secular solutions are given by
\beqa
u_n &=&AP_n(t)e^{-it},\non\\
L(t)P_n(t)&\equiv&\left(\der{^2 }{t^2} -2i \der{ }{t}\right)P_n =
tP_{n-1},\lb{defp}
\eeqa
where $n=1,2,\cdots$ and $P_0=1$,while an initial condition $P_n(0)=0$
is set for $n\ge 1$.
The important observation is that $P_n(t)$ is a polynomial of
degree $2n$ which is given by, for example,
\beqa
P_1&=&(it^2+t)/4,\quad P_2=\sum_{j=1}^4 p_j t^j,\non\\
p_4&=&-1/32,\quad p_3=5i/48,\quad p_2=5/32,\quad p_1=5i/32.\non
\eeqa
Thus, we have
\beq
u=A(1+\eps P_1(t)+\eps^2 P_2(t)+\cdots)e^{-it}+\mbox{c.c.},\non
\eeq
which yields a renormalization transformation $A \to \A(t)$ defined by
the same expression as the previous example \fr{rentr1}, where secular
terms $P_n$ are given by polynomial solutions of \fr{defp}.
Since $\A(t)$ has no longer a translational symmetry in this case,
it is necessary to introduce the following general procedure
in order to derive an asymptotic representation to the Lie group from
\fr{rentr1}.
Let us shift $t \to t+\tau$ in \fr{rentr1}:
\beq
\A(t+\tau)=A(1+\eps P_1(t+\tau)+\eps^2 P_2(t+\tau)+\cdots).\non
\eeq
Noting that $P_n$ is a polynomial of degree $2n$, we have the following
expansion of $\A$ with respect to $\tau$ and $\eps$.
\beqa
\A(t+\tau)&=&A(1+\sum^\infty_{n=1}\eps^n P_n(t))
+[\tau\sum^\infty_{n=1}\eps^nP_{n,t}\non\\
&+&(\tau^2/2)\sum^\infty_{n=1}
\eps^nP_{n,2t}+(\tau^3/3!)\sum^\infty_{n=2}\eps^nP_{n,3t}\non\\
&+&(\tau^4/4!)\sum^\infty_{n=2}\eps^nP_{n,4t}
+\cdots]A,\lb{rentr3}
\eeqa
where $P_{n,kt}$ denotes the k-th derivative with respect to $t$.
Replacing $A$ in \fr{rentr3} with $\A$ by means of the renormalization
transformation \fr{rentr1},  we reach an asymptotic representation to
the Lie group :
\beqa
\A(t+\tau)
&=&\A(t)
+[\tau\sum^\infty_{n=1}\eps^nP_{n,t}+(\tau^2/2)\sum^\infty_{n=1}
\eps^nP_{n,2t}
\non\\
&+&(\tau^3/3!)\sum^\infty_{n=2}\eps^nP_{n,3t}
+(\tau^4/4!)\sum^\infty_{n=2}\eps^nP_{n,4t}\non\\
&+&\cdots]\A(t)/(1+\sum^\infty_{n=1}\eps^n P_n(t)). \lb{lieg1}
\eeqa
Due to loss of the translational symmetry, the right-hand side of
\fr{lieg1} explicitly depends on $t$ and so the representation \fr{lieg1}
should read
\beq
\A(t+\tau)=G_\tau\{\A(t),T(t)\},\quad T(t)=t.\lb{lieg0}
\eeq
From \fr{lieg1} and \fr{lieg0}, we can derive an asymptotic expression of a
 generator of the Lie group, that is, a RG equation:
\beqa
\der{\A}{t}&=&\partial_\tau [G_\tau\{\A(t),T(t)\}]_{\tau=0}\non\\
&=&\der{}{t}\{\ln(1+\sum_{n=1}\eps^n P_n(t))\}\A(t),\lb{lieg2}
\eeqa
from which the following asymptotic solution $\A(t)$ is easily obtained
by integrating a truncated expression of \fr{lieg2}.
\beq
\A(t)=\A(0)\exp\{\eps P_1(t)-\eps^2(P_2(t)+P_1(t)^2)+\cdots\},\lb{solt1}
\eeq
where the argument of $\exp$ function in \fr{solt1} is determined by
integration of a  truncated power series  of
\beq
\der{}{t}(\eps P_1+\eps^2 P_2+\cdots)/(1+\eps P_1+\eps^2 P_2+\cdots).
\lb{epp}
\eeq
The solution \fr{solt1} is a genaralization of the previous result
\cite{cgo1}.
Note that the RG equation \fr{lieg2} is formally integrated to
recover the renormalization transformation \fr{rentr1}.\\

Now, we summarize the above procedures for providing a RG equaion as a
generator of a renormalization group (the Lie group).\\
(1) Caluculate secular terms in a naive perturbation series.\\
(2) Find integral constants $A\in R^n$, which are renormalizable
in order to remove secular terms, and construct a renormalization
transformation $\A(t)=\tilde {R}(A,t)$ which is obtained as a power series
of $t$.\\
(3) Rewrite the renormalization transformation in the form of
the Lie group $\A(t+\tau)=G_\tau\{\A(t),T(t)\}$ by means of
a shift operation $t \to t+\tau$, where $\A$ and $T(t)=t$ constitute
a differentiable manifold on which the Lie-group $G_{\tau}$ acts. \\
(4) A RG equation is obtained as a generator of the Lie group:
$$\der{\A}{t}=\partial_\tau[G_\tau\{\A(t),T(t)\}]_{\tau=0}.$$
If a renormalization transformation $\A(t)=\tilde {R}(A,t)$ has a translational
symmetry with respect to $t$, a repreresentation of the Lie group
does not depend on $T(t)=t$ and takes the following simple form.
\beq
\A(t+\tau)=G_\tau\A(t)=\tilde {R}(\A(t),\tau).\lb{pdtr}
\eeq
In the following subsections, we derive various RG equations
following the above procedures.

\subsection{Mathieu Equation}

As a second example of non-autonomous systems, we derive a RG equation
near the transition curves that separate stable from unstable solutions of
the Mathieu equation
\beq
\der{^2 u}{t^2} + u=-\eps \{\omega+2\cos (2t )\}u,\non
\eeq
where $\omega=\omega_1+\eps \omega_2+\eps^2 \omega_3+\cdots$.
Expanding $u$ just as in the previous example
\beq
u =A e^{-it} +\mbox{c.c.}+\epsilon u_1+\epsilon^2 u_2+\cdots,\non
\eeq
we have, in the leading order,
\beq
\der{^2 u_1}{t^2} + u_1=-(\omega_1A+A^*)e^{-it}-Ae^{-3it}+\mbox{c.c.}, \non
\eeq
of which general solution is
\beq
u_1=-(\omega_1A+A^*)P_1(t)e^{-it}+A/8e^{-3it}+\mbox{c.c.},\non
\eeq
where $A^*$ denotes the complex conjugate of $A$ and $P_1(t)$ is a
polynomial solution of
\beq
\left(\der{^2}{t^2}-2i\der{}{t}\right)P_1\equiv L(t)P_1=1,\lb{lp1}
\eeq
where  $P_1(t)$ is uniquely determined for the initial condition
$P_1(0)=0$, i.e. $P_1(t)=it/2$.
Similar perturbation caluculations yield higher order solutions:
\beqa
u_2&=&-\{(\omega_2+1/8)P_1+(1-\omega_1^2)P_2\}Ae^{-it} \non \\
&+&\{(\omega_1A+A^*)Q_1+\omega_1A/8^2\}e^{-3it}+(A/192)e^{-5it}
+\mbox{c.c.},\non\\
u_3&=&[\{(\omega_1/32-\omega_3)A+(3/64)A^*\}P_1
+\{2(\omega_2+1/8)\non \\
&+&(1-{\omega_1}^2)/2\}\omega_1AP_2+(1-{\omega_1}^2)(\omega_1A+A^*)P_3]
e^{-it}+\mbox{c.c.},\lb{mt3}
\eeqa
where, for simplicity, higher-harmonic terms in \fr{mt3} are not listed
explicitly  and
 secular terms $P_2,\quad P_3,\quad Q_1$ are polynomial solutions of
\beqa
LP_2&=&P_1,\quad LP_3=P^*_2, \non \\
 L_3Q_1&\equiv&\left(\der{^2}{t^2}-6i\der{}{t}-8\right)Q_1=P_1.\lb{lq3}
 \eeqa
Explicit forms of secular terms are given by
\beq
P_2=-t^2/8+it/8,\quad P_3=-it^3/48,\quad Q_1=-it/16-3/64, \lb{secs}
\eeq
where $Q_1$ is a non-resonant secular term in the third higher homonics,
which comes from a resonant secular term $P_1$ in the fundamental harmonic.
In order to remove secular terms in the coefficient of the fundamental
harmonic ($e^{-it}$) to $O(\eps^3)$, we introduce a renormalization
transformation
\beqa
\A(t)&=&\tilde {R}(A,t)\equiv A-\eps(\omega_1A+A^*)P_1
-\eps^2\{(\omega_2+1/8)P_1+(1-\omega_1^2)P_2\}A \non \\
&+&\eps^3[\{(\omega_1/32-\omega_3)A+(3/64)A^*\}P_1
+\{2(\omega_2+1/8)\non \\
&+&(1-{\omega_1}^2)/2\}\omega_1AP_2+(1-{\omega_1}^2)(\omega_1A+A^*)P_3].
\lb{mtrn}
\eeqa
Performing a shift operation $t\to t+\tau$ on \fr{mtrn} and replacing
$A$ in the right-hand side of \fr{mtrn} by $\A(t)+\eps(\omega_1\A+\A^*)
P_1(t)+O(\eps^2)$ , we obtain
a representation of the renormalization group
\beq
\A(t+\tau)=G_\tau\A(t)=\tilde {R}(\A(t),\tau).\lb{mtrn1}
\eeq
This expression indicates that the renormalization group has a translational
symmetry. In fact, \fr{mtrn1} is also directly derived from \fr{mtrn}
when we take into
account of the fact that Fourier components of $u$ obey a coupled but
autonomous system of equations. This result shows an interesting
example where the renormalization group has a translational
symmetry even if the original system does not have one.
It is easy to see that  non-resonant secular terms such as $Q_1$ in
higher harmonics are automatically eliminated when resonant secular terms
in the fundamental harmonic are renormalized by \fr{mtrn}.
Further discussions are presented about automatic elimination of
non-resonant secular terms in higher harmonics in Section {\bf 2.E}.
By differentiating \fr{mtrn1} with respect to $\tau$, we have a generator
of the  Lie group \fr{mtrn1}
\beqa
\der{\A}{t} &=&-(i\eps/2)(\omega_1{\tilde A}+{\tilde A}^*)-(i\eps^2/2)
 \Delta \A \non \\
&+&(i\eps^3/2)[ (\omega_1/32-\omega_3)\A+(3/64)\A^*
+(\omega_1/2)\Delta\A], \lb{mtge}\\
\Delta&=&\omega_2+1/8+(1-\omega_1^2)/4,\non
\eeqa
from which
\beqa
\der{^2 \A}{t^2}&=&[\eps^2(1-\omega_1^2)/4-(\eps^3\omega_1)\Delta \non \\
&+&\eps^4\{\omega_1(\omega_1/32-\omega_3)-3/64+\omega_1\Delta/2\}]\A. \non
\eeqa
which gives  the transition curves that separate stable
from  unstable solutions: $\omega_1=\pm 1, \quad \omega_2=-1/8, \quad
\omega_3=\mp1/64$.
 This transition curves $\omega_1, \omega_2$ and a special RG equation for
$\omega_1=1$ were obtained to $O(\eps^2)$ by means of the RG method
\cite{cgo1}. The expression of a generator \fr{mtge} is a generalization
of the previous results.

\subsection{Large-scale expansion in Gravitational Equation }

A uniformly and isotropically expanding space with dust is discribed by the
following Einstein's equation called the Friedman-Robertson-Walker
equation .
\beq
2a\add+\ad^2=-(k-\Lambda a^2), \lb{frw}
\eeq
where $a$ is a time-dependent scale factor of space, a dot on $a$ denotes
the derivative with respect to time $t$, $k(=\pm 1, 0)$ is a
sign of curvature and
$\Lambda$ is the cosmological constant. Although solutions of \fr{frw} with
 an initial condition $a(0)=0$ are well-known and describe various kinds of
expanding and/or contracting space, this example gives a peculiar
asymptotic form of the representation of the Lie group for
a large-scale solution as is shown later. That is, the higher order
RG equation becomes non-autonomous though the original equation \fr{frw}
and the leading order RG equation are autonomous. \\
The leading order RG equation has been derived for a large-scale solution
of \fr{frw} with $\Lambda=0$ by Y.Nambu and Y.Y.Yamaguchi
\cite{nambu}  and they shows that
a contraction stage of the Friedman space ($k=1$) is qualitatively
recovered by only the leading order RG equation.\\
 Here, we derive a  higher order and non-autonomous RG equation
 by means of the present Lie-group approach for the purpose of
 quantitative comparison with the exact results. \\
 If $\Lambda=0$, we can expand
 $a$ around a large-scale solution $a_0=At^{2/3}$ as
 \beq
 a=x\{A+a_1(x)/A+a_2(x)/A^3+a_3(x)/A^5+\cdots\},\lb{gex1}
 \eeq
where $x=t^{2/3}$, $A$ is an arbitrary constant and $A\gg 1$.
Substituting \fr{gex1} into \fr{frw} and solving the resulting equation
order by order with respect to $A^{-1}$, we
obtain the following secular terms
\beqa
a_j&=&-c_jx^j \quad j=1,2,3,....,\lb{gsec}\\
c_1&=&\frac{9k}{20},\quad c_2=\frac{3c_1^2}{7},\quad
 c_3=\frac{23c_1c_2}{27}, \quad c_4=\frac{9c_2^2+19c_1c_3}{22}.\non
 \eeqa
Renormalizing the integral constant $A$ so that secular terms \fr{gsec}
are removed, we have a renormalization transformation
\beq
\A(x)=A+a_1(x)/A+a_2(x)/A^3+a_3(x)/A^5+\cdots.\lb{grnt}
\eeq
Replacing $x$ in \fr{grnt} by $x+\xi$, we obtain a non-autonomous
representation of the Lie group:
$$\A(x+\xi)=G_{\xi}\{\A(x),x\},$$
of which generator is explicitly given as
\beqa
\der{\A}{x}&=&\partial_{\xi}G_{\xi}(\A(x),x)|_{\xi=0}\non\\
&=&-\frac{c_1}{\A}+\frac{(c_1^2-2c_2)x}{\A^3}
-\frac{(2c_1^3-7c_1c_2+3c_3)x^2}{\A^5}\non\\
&+&\frac{(5c_1^4-24c_1^2c_2+6c_2^2+16c_1c_3-4c_4)x^3}{\A^7}
+\cdots. \lb{ggen}
\eeqa
It is easy to see that this expression of the generator is asymptotic for
$x/\A^2=t/a(t)< 1$, that is, the scale factor $a$ is greater than the
cosmic horizon.
Results of numerical integration of \fr{ggen} for $k=1$ are compared with
the exact solution in Fig.(1). Solutions of the RG equation
 with higher-order corrections \fr{ggen} fits the exact solution
 closer in the region $t/a(t)< 1$ but they separate away from the exact
 solution for  $t/a(t)>1$ where the present asymptotic expansion is not
 valid. It may be interesting that the leading order
 RG solution gives the best fit for $t/a(t)>1$. This may come from
 fact that the leading order RG equation does not have the factor
 $x/\A^2=t/a(t)$ , which should be small in the present asymptotic expansion
 .\\
 
When  a value of $\Lambda$ is near the critical value where the Friedman
space begins to expand again, $\Lambda$ should be scaled as
\beq
\Lambda=\lambda/A_0^6,\non
\eeq
where $A_0=\A(0)$ and the critical value of $\Lambda$ is given
by $\lambda_c=9/4$. Then, effects of this small $\Lambda$ enter into
\fr{ggen} as the following correction terms of $O(\A^{-5})$ and
$O(\A^{-7})$
\beq
3\lambda_3\A x^2+2(c_1\lambda_3-2\lambda_4)x^3,\lb{gcor}
\eeq
where $\lambda_3=\lambda/(12A_0^6), \lambda_4=9\lambda c_1/(88A_0^6)$.
Numerically integrating \fr{ggen} with corrections \fr{gcor}, we obtain
asymptotic solutions near the critical value $\Lambda_c$ as shown in
Fig.(2), which demonstrates  again that asymptotic solutions agree well
 with an exact solution for $t/a(t)<1$.

\begin{figure}[hbp]

\begin{flushleft}
\epsfxsize=14cm
\epsffile{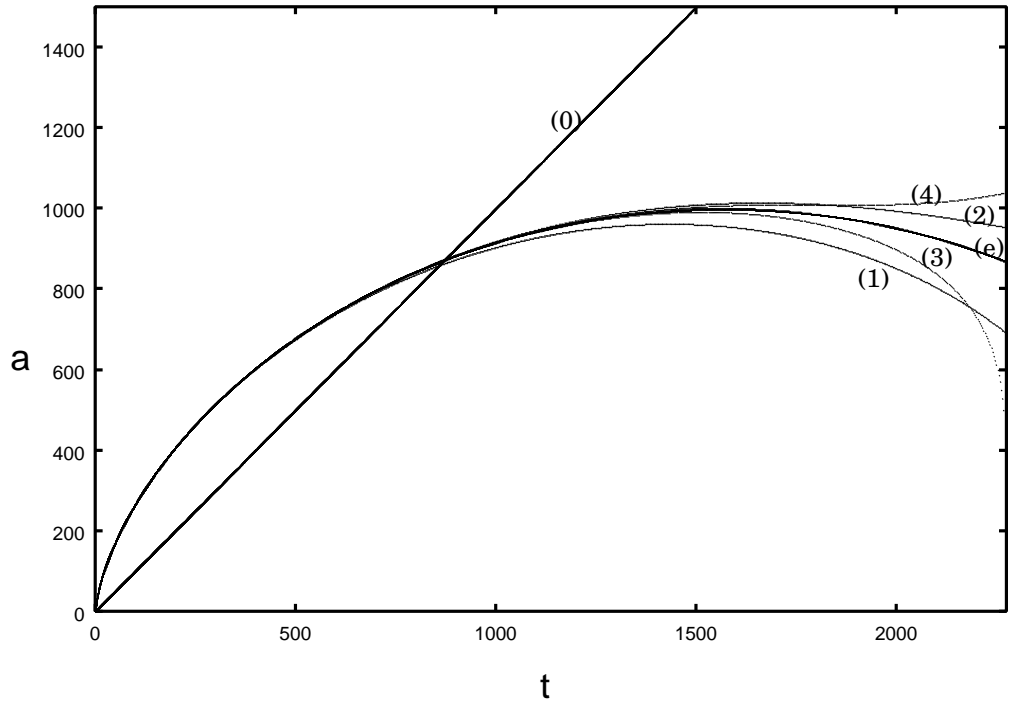}
\end{flushleft}
\caption{
The scale factor $a$ vs. $t$ for $\Lambda=0$ the line 0 is the cosmic horizon,
 i.e. $a=t$, the curve e is the exact solution, the curve 1 is a solution of
the leading order RG equation, the curves 2 ,3 and 4 are solutions of the
second and the third and the fourth order RG equations respectively.}

\end{figure}

\begin{figure}[hbp]
\begin{flushleft}
\epsfxsize=14cm
\epsffile{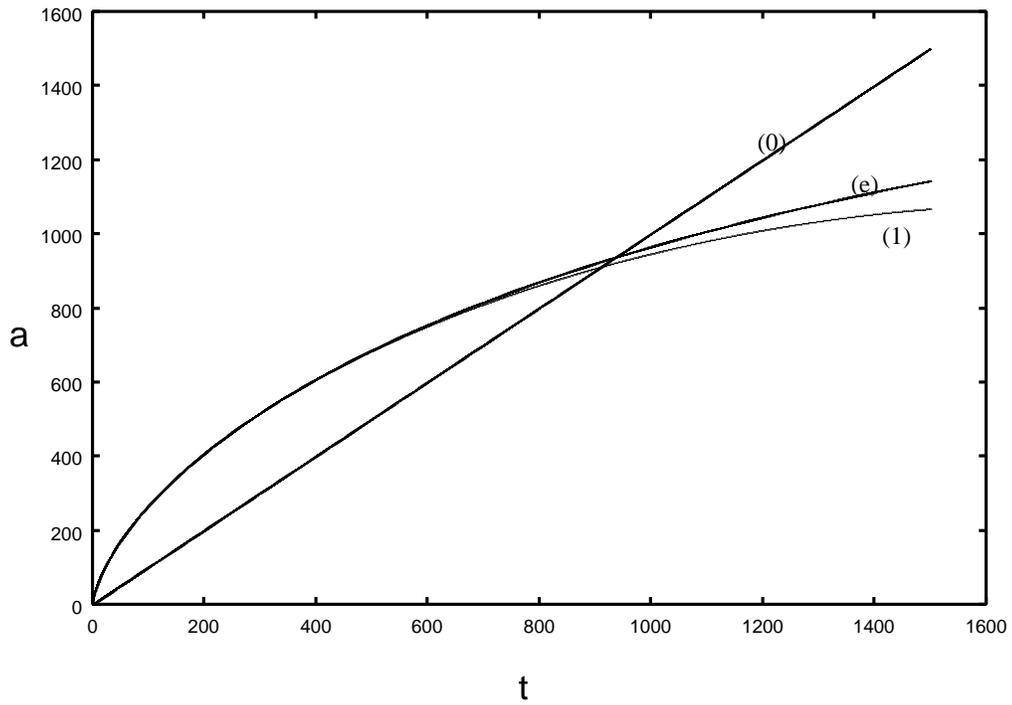}
\end{flushleft}
\caption{
The scale factor $a$ vs. $t$ for $\Lambda=1.1 \Lambda_c$ the line 0 is the
cosmic horizon ($a=t$), the curve e is the exact solution, the curve 1 is a
solution of the RG equation.}
\end{figure}

\pagebreak
\subsection{Weakly Nonlinear Oscillator}

Let us apply  the Lie-group approach to
 the motion of a weakly nonlinear oscillator.
\beq
\der{^2 u}{t^2} + u =\eps a_3 u^3+\eps^2 a_5 u^5+\eps^3 a_7 u^7+\cdots.
\lb{no1}
\eeq
Expanding $u$ as
\beq
u =A e^{-it} + \mbox{c.c.}+\epsilon u_1+\epsilon^2 u_2+\cdots,\non
\eeq
we have
\beq
\left(\der{^2 }{t^2} + 1\right)u_1 = a_3[3|A|^2A e^{-it}
+A^3 e^{-3it}+\mbox{c.c.}]. \non
\eeq
The general solution is given by
\beq
u_1 =a_3[3|A|^2AP_1 e^{-it}+A^3q_1e^{-3it}+\mbox{c.c.}],\non
\eeq
where $q_1=-1/8$ and $P_1(t)$ is a polynomial solution of \fr{lp1}.
For $O(\eps^2)$ , we have
\beqa
u_2&=&\{3a_3^2(3P_2+q_1P_1)+10a_5P_1\}|A|^4A e^{-it}\non\\
&+&\{9a_3^2Q_1+(6a_3^2q_1+5a_2)q_1\}|A|^2A^3e^{-3it}\non\\
&+&(3a_3^2q_1+a_5)q_2A^5e^{-5it}+\mbox{c.c.},\non\\
LP_2&=&P_1,\quad q_2=-1/24,\quad L_3Q_1=P_1,\lb{lp2}
\eeqa
where $L,\quad L_3$ are defined in \fr{lp1} and \fr{lq3}.
Explicit forms of secular terms are given in \fr{secs}.
 Up to $O(\eps^2)$, the secular solution is given by
\beqa
u&=&\left(A+3\eps a_3|A|^2AP_1+\eps^2 \{3a_3^2(3P_2+q_1P_1)
+10a_5P_1\}|A|^4A\right) e^{-it} \non \\
&-&(\eps /8)\{a_3+\eps (9a_3^2P_1+21a_3^2/8+5a_5)|A|^2\}A^3e^{-3it}
+\mbox{c.c.}+\cdots. \lb{ssno}
\eeqa

In \fr{ssno},  coefficients of $e^{-it}$  are
summed up to give  an asymptotic expansion to the
renormalized amplitude $\A$ .
\beqa
\A(t)&=&A+3\eps a_3|A|^2AP_1(t)+\eps^2\{3a_3^2(3P_2(t)+q_1P_1(t))
\non \\
&+&10a_5P_1(t)\}|A|^4A+O(\eps^3),\lb{aa0}
\eeqa
while coefficients of the third harmonic are summed up to give
$\eps\A_3(t)$ as
\beqa
\eps\A_3(t)&=&-(\eps /8)[a_3+9\eps a_3^2|A|^2P_1(t)\non\\
&+&\eps(21a_3^2/8+5a_5)|A|^2+O(\eps^2)]A^3.\lb{aa3}
\eeqa
Substituting an iterative expression of $A$ in terms of $\A$ obtained from
\fr{aa0} into \fr{aa3}, we can eliminate a secular term in \fr{aa3} and
have
\beq
\A_3(t)=-1(/8)\A(t)^3\{a_3+\eps(21a_3^2/8+5a_5)|\A(t)|^2+O(\eps^2)\}.
\lb{aa33}
\eeq
This elimination of secular terms in the higher harmonics is quite
general as also seen in  Section {\bf 2.C}
because secular terms in the higher harmonics originate directly
from  resonant secular terms in the fundamental harmonic and should
disappear as soon as the amplitude $A$ of the fundamental harmonic
is renormalized by the renormalization transformation \fr{aa0}.
In this sense, we call secular terms in the higher harmonics as
non-resonant secular terms, which disappear when resonant secular terms are
 removed by the renormalization transformation.\\
 \noi
Performing a transformation
$t \rightarrow t+\tau$ on \fr{aa0} and
using explicit forms of secular terms , we obtain
\beqa
\A(t+\tau)&=&\A(t)+3\eps a_3|\A(t)|^2\A(t)P_1(\tau)+
\eps^2\{3a_3^2(3P_2(\tau) \non \\
&+&q_1P_1(\tau)) +10a_5P_1(\tau)\}|\A(t)|^4\A(t)+O(\eps^3),
\lb{aren1}
\eeqa
which is an explicit representation of the Lie group underlying
the renormalization transformation \fr{aa0}.
Since the original equation of motion \fr{no1} is translatinally invariant
, the amplitude of osillation $\A$ also enjoyes a translational symmetry.
Thus,  \fr{aren1} is also a direct result of \fr{aa0} and the
translational symmetry.
Since we could proceed this perturbation calculations to arbitrarily
higher order although calculations becomes tedious for the higher order,
 we can also extend \fr{aren1} up to arbitrary order. Noting the
 important observation  that  secular terms appearing in the coefficient
 of the fundamental harmonic $\exp(-it)$ in $u_n$ are
polynomials of degree $n$ and proportinal to $|A|^{2n}A$, we have
\beqa
\A(t+\tau)&=&\A(t)
+\sum^\infty_{n=1}\eps^n\sum^n_{j=1}p_{n,j}\tau^j|\A(t)|^{2n}\A(t) \non\\
&=&\A(t)+\eps p_{1,1}\tau |\A(t)|^2\A(t) \non\\
&+& \eps^2(\tau^2p_{2,2}+\tau p_{2,1})|\A(t)|^4\A(t)+\cdots, \non
\eeqa
where $p_{n,j}$ is constant.
Equating coefficients of equal powers of $\tau$ in \fr{aren1} and
\fr{aulig},  we have a set of  equations
\beq
\der{^n\A}{t^n}/n!=\eps^n \sum^\infty_{j=0}\eps^jp_{n+j,n}
|\A(t)|^{2(n+j)}\A(t).\lb{rge}
\eeq
 Some explicit forms
of $p_{n,j}$ are given as
\beqa
p_{1,1}&=&i(3/2)a_3, \quad p_{2,1}=i\{(15/16)a_3^2+5a_5\},\non\\
 p_{3,1}&=&i\{(123/128)a_3^3+5a_3a_5+(37/2)a_7\} \non \\
p_{2,2}&=&-(9/8)a_3^2,\quad p_{3,2}=-a_3(45a_3^2/32+15a_5/2), \non\\
p_{3,3}&=&-(9i/16)a_3^3, \non
\eeqa
The other $p_{n,j}$ are obtained when higher order secular terms
are calculated. Thus, for $n=1$, eq.\fr{rge} gives an
asymptotic representation for a generator of the Lie group.
\beqa
\der{\A}{t}&=& \sum^\infty_{j=0}\eps^jp_{1+j,1}|\A(t)|^{2(1+j)}\A(t)
\non\\
&=& \eps i(3/2)a_3|\A|^2\A + \eps^2
i\{(15/16)a_3^2+5a_5\}|\A|^4\A+\cdots. \lb{rgee}
\eeqa
Since $p_{1+j,1}$ is purely imaginary, it is easy to see that $|\A|^2$ is
a constant of motion and we formally integrate \fr{rgee} to have an
asymptotic expression to the amplitude of the fundamental harmonic
by all order:
\beq
\A(t)=A\exp[i\sum^\infty_{j=0}\eps^jp_{1+j,1}|A|^{2(1+j)}A t],\non
\eeq
where $A$ is an integral constant. Then, as noted earlier, secular terms
in the higher harmonics are removed automatically, we finally obtain
an asymptotic solution of \fr{no1}
\beq
u=\A(t)e^{-it} +\eps A_3(\A)e^{-3it} +\eps^2 A_5(\A)e^{-5it}+\cdots,
\lb{noas}
\eeq
where amplitudes of higher hamonics slave to $\A$ (e.g. see \fr{aa33}).
For the case that $a_5=a_7=\cdots=0$, the asymptotic solution \fr{noas}
may be interpreted as an asymptotic Fourier
expansin of the Jacobi's elliptic function, which is the exact solution
of \fr{no1}. \\
As far as the final RG equation \fr{rgee} is concerned, the following
convenient derivation is possible without exploiting explicit forms of
secular terms \lb{stno} . Let us operate the linear operator
$L(\tau)$ defined in \fr{lp1} on the both sides of eq.\fr{aren1} and
set $\tau=0$. Then, taking  \fr{lp1} and \fr{lp2} into account, we have
\beq
L(t)\A(t)=3\eps a_3|\A|^2\A +
\eps^2 \{(-3/8)a_3^2+10a_5\}|\A|^4\A.\lb{prono}
\eeq
After $\partial_t^2$ in $L$ is eliminated iteratively , \fr{prono}
reduces to \fr{rgee}. This procedure is valid for the secular solution
\fr{aren1} including secular terms higher order than $O(\eps^2)$
in the present case. In fact, if this procedure is applied to the first
order secular solution, we have
\beq
L\A=3\eps a_3|\A|^2\A,
\eeq
which
is not correct as the first order RG equation since
$\partial_t^2$ in $L$ is of $O(\eps^2)$ (see \fr{rge}).

\subsection{Adiabatic Perturbation for a Soliton}

Although a soliton is a special solution of PDE,  adiabatic variations of
soliton-parameters are described by a system of ODE and we will see that
a corresponding renormalization group has only one parameter.
 Various perturbation methods for solitons have been developed
based on the inverse scattering transform \cite{kau}. However,
 their applications are limited to nearly integrable systems.
 There are some attempts \cite{bon} to develope perturbation methods
 without utilizing the complete
integrability of the unperturbed state, which are applicable to
a pulse solution of non-integrable systems.
Here, in line with these attempts, we derive the well-known adiabatic
equation to soliton-parameters of the K-dV soliton by means of
the RG method.
Let us consider the K-dV equation with a perturbation
\beq
\partial_tu-6u\partial_xu+\partial_x^3u=\eps f(u).\non
\eeq
Expanding $u$ around a soliton solution
\beq
u_0=s(k,z)\equiv -2k^2\mbox{sech}^2(kz),\quad z=x-v(k)t+\xi,\quad
v(k)=4k^2, \lb{soli}
\eeq
we have, for the first order correction $u_1$,
\beq
\{\partial_t+L(z)\}u_1=f(s),\lb{lkv}
\eeq
\beq
L=-v\partial_z-6(s\partial_z+\partial_z s)+\partial_z^3.\non
\eeq
Since $k$ and $\xi$ in \fr{soli} are arbitrary parameters,
zero-eigenfunctions of $L(z)$ with
a bounded boudary condition
are given by $\partial_zs(k,z)\equiv s', \quad \partial_ks(k,z)\equiv
\dot{s}$. In fact, we see that zero-eigenfunctions are degenerate
\cite{ohta} so that
\beq
Ls'=0,\quad L\dot{s}=\dot{v}s', \quad L^2\dot{s}=0, \lb{eigen}
\eeq
where $\dot{v}=\partial_k v$. Here, let us introduce an adjoint
zero-eigenfunction $\hat{s}$ defined by the adjoint operator $\hat{L}$ to
 $L$
as
\beq
\hat{L}\hat{s}=0, \quad \hat{s}=\mbox{sech}^2(kz). \lb{ajoit}
\eeq
From \fr{eigen} and \fr{ajoit}, we find
\beq
< \hat{s}\cdot s'>\equiv \int_{-\infty}^\infty \hat{s}s'dz=0.\non
\eeq
In terms of these zero-eigenfunctions, we put a spatially bounded
solution as
\beq
u_1=P_1(t)s'+Q_1(t)\dot{s}+\bar{u}_1(z).\lb{sbs}
\eeq
Substitutiing \fr{sbs} into \fr{lkv} , we have an equation for
$\bar{u_1}(z)$
\beq
L(z)\bar{u_1}+(P_{1,t}+\dot{v}Q_1)s'+Q_{1,t}\dot{s}=f(s),\lb{lkv2}
\eeq
where the suffix $t$ denotes the derivative with respect to $t$.
Since \fr{lkv2} is an equation for  $t$
independent and spatially bounded $\bar{u_1}$, the following consistency
conditions must be satisfied:
\beqa
Q_{1,t}&=&\frac{<\hat{s}\cdot f(s)>}{<\hat{s}\cdot \dot{s}>}
=\mbox{constant},\lb{con2}\\
P_{1,t}+\dot{v}Q_1&=&\mbox{constant},\non
\eeqa
from which
\beq
P_{1,2t}=-\dot{v}Q_{1,t}.\lb{con3}
\lb{con1}
\eeq
From conditions \fr{con2} and \fr{con3}, secular terms $P_1(t)$ and
$Q_1(t)$ are found to be polynomials of degree 2 and 1 respectively.
\beq
P_1=p_{1,2}t^2+p_{1,1}t,\quad Q_1=q_{1,1}t,\non
\eeq
where coefficients $p, q$ are
\beq
2p_{1,2}=-\dot{v}q_{1,1},\quad q_{1,1}=\frac{<\hat{s}\cdot f(s)>}
{<\hat{s}\cdot \dot{s}>},\lb{coeff}
\eeq
while $p_{1,1}$ is  an arbitrary constant. \\
Thus, a secular solution to $O(\eps)$ is found to be
\beq
u=s(k,x+\theta)+\eps(P_1(t)s'+Q_1(t)\dot{s}), \non
\eeq
where $\theta=-vt+\xi$.
In order to eliminate  secular terms $P_1, Q_1$  by
renormalizing arbitrary soliton-parameters $k$ and $\xi$,
we introduce a renormalized soliton solution as
\beq
s(\tilde{k}(t),x+\tilde{\theta}(t))=s(k,x+\theta)+\eps\{P_1(t)s'(k,x+\theta)
+Q_1(t)\dot{s}(k,x+\theta) \},\lb{resol}
\eeq
where $\tilde{\theta}(t)=-v(\tilde{k}(t))t+\tilde {\xi}(t)$.
Setting $\tilde{k}(t)=k+\eps k_1(t), \quad \tilde{\theta}(t)=\theta
+\eps \theta_1(t)$, the left-hand side of \fr{resol} is expanded as
\beq
s(\tilde{k},x+\tilde{\theta})=s(k,x+\theta)+\eps(\theta_1s'
+k_1\dot{s} )+O(\eps^2).\lb{resol2}
\eeq
From \fr{resol} and \fr{resol2}, we have a renormalization transformation
\beq
\tilde{k}(t)=k+\eps Q_1(t;k), \quad \tilde{\theta}(t)=\theta+\eps P_1(t;k),
\lb{kvren}
\eeq
where $P_1, Q_1$ depend on $k$ through their coefficient $p, q$
 given in \fr{coeff}. Shifting $t$ to $t+\tau$ in \fr{kvren} and
 replacing $k$
 by $\tilde{k}(t)-\eps Q_1(t;k)$, we obtain a renormalization
 group
\beqa
\tilde{k}(t+\tau)=\tilde{k}(t)+\eps Q_1(\tau;\tilde{k}(t)),\non\\
\tilde{\theta}(t+\tau)=\tilde{\theta}(t)-v(\tilde{k}(t))\tau+
\eps P_1(\tau;\tilde{k}(t)),\non
\eeqa
from which we obtain a generator of the renormalization group
\beqa
\der{\tilde{k}}{t} &=&\eps q_{1,1}(\tilde{k}),\lb{gkv1}\\
\der{\tilde{\theta}}{t}&=&-v(\tilde{k})+\eps p_{1,1}(\tilde{k}),
\lb{gkv2}\\
\der{^2\tilde{\theta}}{t^2}&=&2\eps p_{1,2}(\tilde{k}),\lb{gkv3}
\eeqa
where consistency between \fr{gkv2} and \fr{gkv3} are assured by
an arbitrary coefficient $p_{1,1}$.
Finally, from \fr{coeff}, \fr{gkv1} and \fr{gkv3}, we reach the
well-known adiabatic equations to soliton-parameters to $O(\eps)$
\cite{kar}.
\beqa
\der{\tilde{k}}{t}&=&-\frac{\eps}{4}\int_{-\infty}^\infty
\mbox{sech}^2(\tilde{k}z)f[s(\tilde{k},\tilde{k}z)]dz,\non\\
\der{^2\tilde{\theta}}{t^2}&=&-\dot{v}(\tilde{k})\der{\tilde{k}}{t}=
-\der{v(\tilde{k}(t))}{t}.\non
\eeqa

\setcounter{equation}{0}
\section{Reduction of Partial Differential Equations}
\subsection{ General Discussion}

If we try to extend the Lie-group approach discussed in the preceding
sections  to PDE, there is  a
difficulty to calculate secular terms so that
infinitely many divergent terms such as
 exponentially divergent terms as well as polynomials appear even in a
 perturbed solution of the leading order.  This is due to the fact that
 the dimension of "kernels" without any boundary conditions of a
 linearized partial differential operator is infinite in general.
 However, each divergent solution belonging to "kernels" is accompanied by
 an integral constant and so unsuitable divergent solutions such as
 exponentially divergent solutions
 are eliminated by nullifying corresponding integral constants.
 Thus, we restrict renormalizable secular terms to polynomials even in PDE.
  This restriction is quite natural in the Lie-group appoach since
  a representation of the Lie group with multiple parameters should be
   expanded in terms of a generator as
\beqa
  \A(t+\tau,x+\xi)&=&\exp(\tau\partial_t+\xi\partial_x)\A(t,x)\non\\
  &=&\{1+\tau\partial_t+\xi\partial_x+(\tau\partial_t+\xi\partial_x)^2/2
  +\cdots\}\A(t,x),\lb{pdg}
\eeqa
which is a polynomial with respect to $\tau, \xi$ when the expansion is
truncated.
Even if  secular terms are ristricted to polynomials, there are still
 infinitely many secular solutions in kernels of  a
 linearized partial differential operator. For the purpose of  clasifying
 infinitely many polynomial terms, the following observation is crucial.
In the expansion \fr{pdg}, each monomial is accompnied by  a diffential
operator of the same order, for example, $\xi$ by $\partial_x$, $\tau^2$
by  $\partial_t^2$ and $\xi\tau$ by $\partial_x\partial_t$ etc., while
each differential operator, say, $\partial_t$ maps $\A$ to
the smaller value $\partial_t\A \ll |\A|$ in the autonomous case, which
will be discussed in the following sections.
In this sense, each monomial in
a representation of the Lie group is considered to have the same order of
magnitude as the corresponding differential operator.
On the other hand, for systems with a translational symmetry,
each monomial with respect to
$t, x$ in secular terms in a perturbation series is replaced by
the coresponding monomial with respect to $\tau, \xi$ in a
representation of the Lie group
(see, \fr{pdtr}). Therefore, each monomial
in  secular terms has also the same order of magnitude as the corresponding
differential operator. Thus, monomials in secular terms are
ordered as
\beq
x\gg x^2 \gg x^3,\quad t\gg tx \gg tx^2 ,\quad tx \gg t^2x, \mbox{etc.}.
\non
\eeq
With this kind of order for monomials in mind, we choose  polynomial
 kernels from the lower order by nullifying integral constants accompanying
 higher-order kernels. In particular,
 the leading order secular term consists of the lowest order polynomial,
 i.e. a polynomial of degree one. As order of perturbed solutions increases,
 the higher order polynomial kernels are taken into secular solutions order
 by order so that order of differntial operators accompanied by polynomial
 secular terms is kept consistent.
 Through this procedure, we can choose suitable kernels as necessary
 ingredients of a renormalization transformation. \\
 On the other hand, there is inevitable flexibility appearing in the
 choice of initial setting of perturbation scheme, which brings about
 different RG equations from each other depending on the choice, although
 only one of them is often interesting as shown in the next example.

\subsection{Nonlinear Schr\"odinger Equation}

As the first example, we consider a nonlinear wave equation, which is
just a wave-equation version of the equation for a nonlinear oscillator
\fr{no1}.
\beq
\partial_t^2u-\nabla^2u+u=\eps ^m a u^3, \lb{nw1}
\eeq
where $m$ is a positive integer.
Let us expand $u$ around a plane-wave solution
\beq
u_0=A\exp\{i(kx-\omega t)\}+\mbox{c.c.}, \quad \omega^2=1+k^2, \non
\eeq
as $u=u_0+\eps u_1+\eps u_2+\cdots$. \\
For $m=1$, we have
\beq
\partial_t^2u_1-\nabla^2u_1+u_1=a u_0^3. \lb{nw2}
\eeq
A secular solution of \fr{nw2} is given by
\beqa
u_1&=&3a|A|^2AP_1(t,x,\rvp)e^{i\theta}-(a/8)A^3e^{3i\theta}+\mbox{c.c.},
\lb{nwsl}\\
LP_1&\equiv&\{-2i\omega(\partial_t+\dot{\omega}\partial_x)
+\partial_t^2-\nabla^2\}P_1=1,\lb{nwsc1}
\eeqa
where $\theta=kx-\omega t,\quad \rvp=(y,z),\quad
\dot{\omega}=\partial_k \omega=k/\omega$.
As discussed in Section {\bf 3.A}, it is assumed that the
leading order secular term is a polynomial of degree one, that is
\beq
P_1=p_{1,0,0}t+p_{0,1,0}x+\pv_{0,0,1}\cdot\rvp.\lb{nwp}
\eeq
Substituting \fr{nwp}  into \fr{nwsc1}, we have
\beq
-2i\omega (p_{1,0,0}+\dot{\omega}p_{0,1,0})=1. \lb{nwp1}
\eeq
The secular term is eliminated by means of a renormalization transformation
\beq
\A(t,x,\rvp)=A+ 3\eps a|A|^2AP_1(t,x,\rvp),\non
\eeq
which is rewritten as,
by excuting arbitrary shifts on independent variables,
\beq
\A(t+\tau,x+\xi,\rvp+\vec{\eta})=\A(t,x,\rvp)+
3\eps a|\A|^2\A(t,x,\rvp)P_1(\tau,\xi,\vec{\eta};t,x,\rvp), \lb{nwlie}
\eeq
which gives a representation of the Lie group with multiple parameters to
$O(\eps)$.
Here, it should be noted that the coefficients of a polynomial $P_1$ may
depend on an "initial position", that is  $p_{j,k,l}$ is replaced by
$\p_{j,k,l}(t,x,\rvp)$ in \fr{nwlie}.
It is seen later in \fr{nwt}, \fr{nwx} and \fr{nwr} that this dependence
comes through $\A$.
In terms of a generator of the Lie group, \fr{nwlie} is expanded:
\beqa
\A(t+\tau,x+\xi,\rvp+\vec{\eta})&=&\exp(\tau\partial_t+\xi\partial_x
+\vec{\eta}\cdot\nabla_{\perp})\A(t,x,\rvp)\non\\
&=&(1+\tau\partial_t+\xi\partial_x
+\vec{\eta}\cdot\nabla_{\perp}+\cdots)\non\\ 
&& \mbox{ }  \cdot\A(t,x,\rvp),\lb{nwlix}
\eeqa
where $\nabla_{\perp}=(\partial_y,\partial_z)$.
Equating coefficients of $\tau, \xi, \vec{\eta}$ in \fr{nwlie} and
\fr{nwlix}, we obtain
\beqa
\partial_t\A&=&3\eps a|\A|^2\A\p_{1,0,0},\lb{nwt}\\
\partial_x\A&=&3\eps a|\A|^2\A\p_{0,1,0},\lb{nwx}\\
\nabla_{\perp}\A&=&3\eps a|\A|^2\A\pvt_{0,0,1}.\lb{nwr}
\eeqa
From \fr{nwp1},\fr{nwt} and \fr{nwx}, $p_{j,k,l}$ or $\p_{j,k,l}$  are
eliminated to yield a RG equation
\beq
(\partial_t+\dot{\omega}\partial_x)\A=3i\eps a|\A|^2\A /(2 \omega),\non
\eeq
which gives a steady propagation of the modulation of plane wave under
effects of nonlinearity. \\
In order to include effects of dispersion, we set $m=2$ in the initial
setting of perturbation. Then, instead of \fr{nw2}, we have
\beq
\partial_t^2u_1-\nabla^2u_1+u_1=0, \non
\eeq
of which secular solution is
\beq
u_1=P_0(t,x,\rvp)e^{i\theta}+\mbox{c.c.},\lb{nwsld}
\eeq
where $P_0$ is the lowest order secular solution of $LP_0=0$ given
by the same form as $P_1$ in \fr{nwp} with a different constraint
\beq
p_{1,0,0}+\dot{\omega}p_{0,1,0}=0. \lb{nwp0}
\eeq
If  secular terms in \fr{nwsld} are removed by a similar renormalization
transformation as above, the constraint \fr{nwp0} would yields a RG
equation
\beq
(\partial_t+\dot{\omega}\partial_x)\A=0,\non
\eeq
of which general solution is written as
$\A=\A(x',\rvp), x'=x-\dot{\omega}t$. Therefore, it is convenient to
 introduce a Galilean transformation
\beq
x'=x-\dot{\omega}t, \quad t=t.\non
\eeq
Then, the operator $L$ in \fr{nwsc1} is rewritten and  $LP_0=0$ becomes
\beq
LP_0=\{-2i\omega\partial_t+(\dot{\omega}^2-1)\partial_{x'}^2-
\nabla_{\perp}^2
-2\dot{\omega}\partial_t\partial_{x'}+\partial_t^2\}P_0=0.\lb{nwlop}
\eeq
The lowest order secular solution of \fr{nwlop} is
\beq
P_0=p^{(1)}_{0,1,0} x'+\pv^{(1)}_{0,0,1}\cdot\rvp,\lb{nwdsc0}
\eeq
where $p^{(1)}_{0,1,0}$ and $\pv^{(1)}_{0,0,1}$ are arbitrary
constants.
The second order secular solution $u_2$ is given by the same form as $u_1$
in \fr{nwsl}, where $P_1$ is the second order polynomial solution of
$LP_1(t,x',\rvp)=1$, that is
\beqa
P_1&=&p^{(2)}_{0,2,0}x'^2+p^{(2)}_{0,1,0}x'
+\pv^{(2)}_{0,0,2}\cdot\rvp\rvp \non \\
&+&\pv^{(2)}_{0,0,1}\cdot\rvp+p^{(2)}_{1,0,0}t,\lb{nwdsc1}\\
-2i\omega p^{(2)}_{1,0,0}&+&2(\dot{\omega}^2-1)p^{(2)}_{0,2,0}
-2\mbox{Tr}[\pv^{(2)}_{0,0,2}]=1, \lb{nwdc}
\eeqa
where $\mbox{Tr}[\pv^{(2)}_{0,0,2}]$ denotes the trace of a matrix
$\pv^{(2)}_{0,0,2}$.
In terms of secular terms \fr{nwdsc0} and \fr{nwdsc1}, a secular solution
to $O(\eps^2)$ is obtained as
\beq
u=(A+\eps P_0+3\eps^2 a|A|^2AP_1)e^{i\theta}-\eps^2(a/8)A^3e^{3i\theta}
+\mbox{c.c.},\non
\eeq
which yields a renormalization transformation
\beq
\A(t,x',\rvp)=A+\eps P_0+\eps^2 3a|A|^2AP_1.\lb{nwrt}
\eeq
Performing the shift operation $t \to t+\tau, x' \to x'+\xi, \rvp \to
\rvp+\vec{\eta}$ , we can rewrite \fr{nwrt} to obtain a representation
of the Lie group
\beqa
\A(t+\tau,x'+\xi,\rvp+\vec{\eta})&=&\A(t,x',\rvp)+
\eps P_0(\tau,\xi,\vec{\eta};t,x',\rvp)\non\\
&+&\eps^2 3a|\A|^2\A(t,x',\rvp)P_1(\tau,\xi,\vec{\eta};t,x',\rvp)
.\lb{nwdlie}
\eeqa
In deriving \fr{nwdlie}, we have used
arbitrariness of the coefficients of $P_0$.
Again, note that coefficients of secular terms $P_j$ depend on an "initial
position" $(t,x',\rvp)$.  As in the preceding
example, this dependence comes through $\A$ and so the Lie group
\fr{nwdlie} also enjoys
 a translational symmetry which the original system \fr{nw1} has.
 Expanding the right-hand side of \fr{nwdlie} in terms of the generator in
 the same form as \fr{nwlix}, we have
\beqa
\partial_{x'}\A&=&\eps \p^{(1)}_{0,1,0}+\eps^2 \p^{(2)}_{0,1,0},\lb{nwax}\\
\nabla_{\perp}\A&=&\eps\pvt^{(1)}_{0,0,1}+\eps^2\pvt ^{(2)}_{0,0,1},
\lb{nwar}\\
\partial^2_{x'}\A&=&6\eps^2 a|\A|^2\A \p^{(2)}_{0,1,0},\lb{nwx2}\\
\nabla^2_{\perp}\A&=&6\eps^2 a|\A|^2\A \mbox{Tr}[\pvt^{(2)}_{0,0,2}],
\lb{nwr2}\\
\partial_t\A&=&3\eps^2 a|\A|^2\A \p^{(2)}_{1,0,0}.\lb{nwt1}
\eeqa
While \fr{nwax} and \fr{nwar} give order of operators $\partial_{x'}\sim
\nabla_{\perp}\sim O(\eps)$,  they do not put any constraints on $\A$
because $\p^{(1)}_{0,1,0}$ and $\pvt^{(1)}_{0,0,1}$ are arbitrary
coefficients of kernels of the linerlized operator $L$. On the other hand,
\fr{nwx2}, \fr{nwr2}, \fr{nwt1} and the constraint \fr{nwdc} yield
the following nonlinear Schr\"odinger equation as a RG equation
\beq
i\partial_t\A+(\ddot{\omega}/2)\partial^2_{x'}\A
+1/(2\omega)\nabla^2_{\perp}\A
+3\eps^2 /(2\omega)a|\A|^2\A=0,\lb{nsch}
\eeq
where $\ddot{\omega}=\partial_k^2\omega=(1-\dot{\omega}^2)/(2\omega)$.\\
It is possible to derive the final result \fr{nsch} formally without
exploiting explicit forms of secular terms by
operating the linearized operator $L(\tau,\xi,\vec{\eta})$ defined in
\fr{nwlop} on both sides of \fr{nwdlie}.  Noting that $LP_0=0, LP_1=1$ and
discarding $\partial_t^2\A, \partial_t\partial_{x'}\A$ as higher order
terms, we immediately obtain the nonlinear Schr\"odinger equation \fr{nsch}.

\subsection{Kadomtsev-Pitviashvili-Boussinesq Equation }

Let us consider a  weakly dispersive nonlinear wave equation:
\beq
\partial^2_t u-\nabla\cdot\{a(u)\nabla u-\eps^2
 b(u)\nabla(\nabla^2 u)\} = 0,\non
 \lb{weakdisp}
\eeq
where the coefficient($\eps^2 b(u)$) of a dispersive term is assumed to
be as small as $\eps^2$.
Expanding $u$ around  constant $u_0$ as
\beqa
u &= &u_0 + \eps (\eps u_1 + \eps^2 u_2 + \eps^3 u_3+\eps^4 u_4+\cdots),
\non\\
a(u)&=&a_o+\eps[a'_0(\eps u_1+\eps^2 u_2)+\eps^3\{a'_0 u_3+(a''_0/2)u_1^2\}
+ \cdots],\non\\
b(u)&=&b_0+\eps b'_0(\eps u_1+\eps^2 u_2) +\cdots,\non
\eeqa
where a dash denotes the derivative with respect to $u$ and
$a_0=a(u_0)=v^2>0$ is assumed. The above setting of perturbation is
justified
a posteriori.
We have the leading-order perturbed  equation
\beq
 \partial_t^2u_1-v^2\nabla^2 u_1\equiv L u_1= 0.\non
\eeq
We may write a simple-wave solution of this linear wave equation in terms of
an arbitrary function $h$:
\beq
u_1 =h(\xi),\non
\eeq
where $\xi=x-vt$. Here, it is convenient to rewrite the linear operator
$L$  by means of
a Galilean transformation $\xi=x-vt,\quad t=t$ as
\beq
L(t,\rvp;\xi)=\partial_t^2 -2v\partial_t\partial_\xi-v^2\nabla^2_{\perp},
\non
\eeq
where $\rvp =(y,z), \nabla^2_{\perp}=\partial^2_y+\partial^2_z$.
The next order correction obeys
\beq
Lu_2= 0,\lb{kp2}
\eeq
whose lowest order secular solution with respect to $t, \rvp$ is given as
\beq
u_2(t,\rvp;\xi)=\pv_{0,1}^{(0)}(\xi)\cdot\rvp,\lb{kpu2}
\eeq
where a monomial of $t$ is discarded because its coefficient is constant.
>From the next order equation, we have
\beq
Lu_3= f_\xi,\qquad f=a'_0hh_\xi-b_0h_{3\xi},\lb{kp3}
\eeq
where the subscript $\xi$ of $h, f$ denotes the partial derivative with
respect to $\xi$. Since  forcing terms in the right-hand side of
\fr{kp3} depend only on $\xi$, \fr{kp3} has a  non-trivial secular
solution with respect to $t, \rvp$:
\beqa
u_3&=&\pv^{(1)}_{0,1}(\xi)\cdot\rvp+\pv^{(1)}_{0,2}(\xi)\cdot\rvp\rvp
+p_{1,0}^{(1)}(\xi)t,\non\\
f_\xi&=&-2v\partial_\xi p_{1,0}^{(1)}(\xi)
-v^2\mbox{Tr}[\pv^{(1)}_{0,2}(\xi)],\lb{kpcon1}
\eeqa
where $\pv^{(1)}_{0,2}(\xi)$ is a $2\times 2$ marrix function of $\xi$.
Summing up secular terms up to $O(\eps^4)$, we obtain
\beq
(u-u_0)/\epsilon^2=h+\epsilon u_2+\epsilon^2 u_3.\non
\eeq
In order to remove secular terms with respect to $t,\rvp$ in $u_2, u_3$
by renormalizing an arbitrary function $h$, we introduce the following
renormalization transformation
\beq
\h(t,\rvp;\xi)=h(\xi)+\epsilon u_2(t,\rvp;\xi)+\epsilon^2 u_3(t,\rvp;\xi).
\lb{kprt}
\eeq
Since $\h(t,\rvp;\xi)$  has a translational symmetry with respect to
$t,\rvp$ as well as $u(t,\rvp;\xi)$,
\fr{kprt} immediately leads to a representation of the Lie group with
arbitrary parameters $\tau, \etav$
\beq
\h(t+\tau,\rvp+\etav;\xi)=h(t,\rvp;\xi)+\epsilon u_2(\tau,\etav;\xi)
+\epsilon^2 u_3(\tau,\etav;\xi),
\lb{kplie1}
\eeq
where it is noted again that coefficients of secular terms $p$ depend on
an "initial position", e.g. $\pv^{(1)}_{0,2}(\xi)$ is replaced by
$\pvt^{(1)}_{0,2}(t,\rvp;\xi)$.
Expanding the right-hand side of \fr{kplie1} in terms of a generator
$\tau\partial_t+\etav\cdot\nabla_{\perp}$ and equating the same
monomials , we have
\beqa
\nabla_{\perp}\h&=&\eps\pvt^{(0)}_{0,1}(t,\rvp;\xi)+
\eps^2\pvt^{(1)}_{0,1}(t,\rvp;\xi),\lb{kph1}\\
\nabla^2_{\perp}\h&=&\eps^2\mbox{Tr}[\pvt^{(1)}_{0,2}(t,\rvp;\xi)],\lb{kph2}\\
\partial_t\h&=&\eps^2\p^{(1)}_{1,0}(t,\rvp;\xi).\lb{kpht}
\eeqa
Since $\pvt^{(j)}_{0,1}$ are arbitrary coefficients of kernels of the
linearized operator $L$, \fr{kph1} does not
put any constraints on $\h$ but provides order of the operator
 $\nabla_{\perp} \sim O(\eps)$.
From \fr{kph2}, \fr{kpht} and the constraint \fr{kpcon1}, we obtain
the Kadomtsev-Pitviashvili (K-P) equation in the three dimensional space
\beq
2v\partial_t\h_\xi+v^2\nabla^2_{\perp}\h+\eps^2
(a'_0\h\h_\xi-b_0\h_{3\xi})=0.\non
\eeq
Next, we calculate higher-order corrections to the K-P equation.
For $u_4, u_5$, we have
\beqa
Lu_4&=&\partial^2_\xi(a'_0hu_2-b_0\partial^2_\xi u_2),\lb{kp4}\\
Lu_5&=&(a''_0/2)(h^2h_\xi)_\xi+a'_0h\nabla^2_{\perp}u_3
+a'_0|\nabla_{\perp}u_2|^2 \non\\
&-&b'_0(hh_{3\xi})_\xi-2b_0\partial^2_\xi\nabla^2_{\perp}u_3
+g(t,\rvp;\xi), \lb{kp5}
\eeqa
where $g(0,0;\xi)=0$. Explict secular solutions of \fr{kp4} and \fr{kp5}
are not necessary for later discussions but we note
\beq
 [Lu_4]_{\rvp=0}=0. \lb{kp41}
\eeq
Then, we obtain a representation of the Lie group with higher order
corrections
\beqa
\h(t+\tau,\rvp+\etav;\xi)&=&h(t,\rvp;\xi)+
\epsilon u_2(\tau,\etav;\xi)
+\epsilon^2 u_3(\tau,\etav;\xi)\non \\
&+&\epsilon^3 u_4(\tau,\etav;\xi)
+\epsilon^4 u_5(\tau,\etav;\xi).
\lb{kplie2}
\eeqa
Since it is straightfoward but tedious to derive the K-P equation with
 higher order corrections from \fr{kplie2} by using explict forms of
 secular solutions, we follow the simplified procedure mentioned in the
 last paragraph in the section {\bf 3.B}.
Operating  $L(\tau,\etav;\xi)$ on both sides of \fr{kplie2} at
$(\tau,\etav)=(0,0)$ and noting \fr{kp2} , \fr{kp3}, \fr{kp41} and
\fr{kp5}, we  obtain  the following RG equation
\beqa
& &(\partial_t^2-2v\partial_t\partial_\xi)h-\nabla_{\perp}\{
(a_0+\eps^2 a'_0h)\nabla_{\perp}h\} \nonumber \\
&=&\eps^2\partial_\xi\{(a'_0h+\eps^2 a''_0h^2/2)h_\xi
-(b_0+\eps^2 b'_0h)h_{3\xi}-2\eps^2b_0\nabla^2_{\perp}h_\xi\}.\lb{hkp}
\eeqa
If $\nabla_{\perp}=0$ ,\fr{hkp} reduces to the Boussinesq equation with
higher order corrections.
 Therefore, \fr{hkp} may be considered as not only the K-P equation
with higher order corrections but also the Boussinesq equation including
multi-dimensional effects and  higer order corrections.

\subsection{Phase Equations }

First, we consider interface dynamics in the following genaral
reaction-diffusion system.
\beq
\partial_t U=F(U)+D\nabla^2U,\lb{rdif1}
\eeq
where $U$ is an $n$-dimensional vector and $D$ is a $n\times n$ constant
matrix. Suppose \fr{rdif1} has an interface solution $U_0=U_0(x-vt+\phi)$,
where $\phi$ is an arbitrary constant, then
\beq
vU_{0,\theta}+F(U_0)+DU_{0,2\theta}=0, \lb{rdif2}
\eeq
where $\theta=x-vt+\phi$ and the suffix $\theta$ denotes the derivative
with rspect to $\theta$. Differentiating \fr{rdif2}, we have
\beqa
L(\theta)U_{0,\theta}&\equiv&-(v\partial_\theta+F'(u_0)\cdot
+D\partial^2_\theta)U_{0,\theta}=0,
\lb{rdif3}\\
LU_{0,2\theta}&=&F''\cdot U^2_{0,\theta},\lb{rdif4}
\eeqa
where $F'\cdot V=(V\cdot \nabla_{U})F(U)|_{U=U_0},\quad F''\cdot V^2=
(V\cdot \nabla_{U})^2F(U)|_{U=U_0}$.
It is convenient to introduce the following coordinate transformation
\beq
\theta=x-vt+\phi,\quad x'=x, t'=t.\non
\eeq
Then, we have
\beqa
\partial_t&=&\partial_{t'}-v\partial_\theta,\quad
 \partial_x=\partial_{x'}+\partial_\theta,\non\\
  \nabla^2&=&\partial^2_\theta+2\partial_\theta\partial_{x'}
  +\nabla^2_{x'},\non
\eeqa
where $\nabla^2_{x'}=\partial^2_{x'}+ \nabla^2_{\perp}, \nabla_{\perp}=
(0,\partial_y,\partial_z)$.
We study the solution close to $U_0$ so that $U$ is expanded as
\beq
U=U_0+\eps U_1+\eps^2 U_2+\cdots.\non
\eeq
The first order term obeys
\beq
\{\partial_{t'}+L(\theta)-(\nabla^2_{x'}+2\partial_\theta\partial_{x'})D\}
U_1\equiv \tilde{L}U_1=0, \lb{inu1}
\eeq
The lowest order secular solution of \fr{inu1} is supposed to be
\beq
U_1=P_1(t',\rv)U_{0,\theta}(\theta),\non
\eeq
where $P_1$ is a scalar secular function of $t'$ and  $\rv=(x',y,z)$.
Then, we have
 $$\partial_{t'}P_1=0,\quad \partial_{x'}P_1=0, \quad
\nabla^2_{x'}P_1=0,$$
from which
\beq
P_1=\nabla_{\perp}P_1\cdot \rvp,\lb{insc1}
\eeq
where $\rvp=(0,y,z)$ and $\nabla_{\perp}P_1$ is a constant
three-dimensional vector.  \\
The second order term $U_2$ is governed by
\beq
\tilde{L}U_2=(1/2)F''\cdot U_1^2, \lb{inu2}
\eeq
of which secular solution is assumed to be
\beq
U_2=\bar{U}_2(\theta)+P_2(t',\rv)U_{0,\theta}+R_2(t',\rv)U_{0,2\theta},
\lb{inu2s}
\eeq
where $\bar{U}_2$ is a bounded function of $\theta$ and $P_2, R_2$ are
scalar secular terms.
Substituting \fr{inu2s} into \fr{inu2} and noting \fr{rdif3} and
\fr{rdif4}, we obtain
\beqa
& &L(\theta)\bar{U}_2+\partial_{t'}P_2U_{0,\theta}
-\nabla^2_{x'}P_2DU_{0,\theta}-2\partial_{x'}P_2DU_{0,2\theta}\non\\
&+&\partial_{t'}R_2U_{0,2\theta}
-\nabla^2_{x'}R_2DU_{0,2\theta}-2\partial_{x'}R_2DU_{0,3\theta}\non\\
&+&\{R_2-(1/2) P_1^2\}F''\cdot U_{0,\theta}^2=0,\non
\eeqa
which is a equation for $\bar{U}_2$. Since $\bar{U}_2$ is a
function of $\theta$ only, the following  conditions
are necessary.
\beq
\partial_{t'}P_2=c_1,\quad \partial_{x'}P_2=c_2,
\quad \nabla^2_{\perp}P_2=c_3,\quad
R_2=(1/2) P_1^2+c_4,\lb{intr}
\eeq
where $c_j$ are constant. Since $P_1$ is a linear function of $\rvp$
given in \fr{insc1}, \fr{intr} yields
\beq
\partial_{t'}R_2=\partial_{x'}R_2=0,\quad
\nabla^2_{x'}R_2=|\nabla_{\perp}P_1|^2.\non
\eeq
Then, a bounded solution $\bar{U}_2$ is possible only if the following
compatibility condition is satisfied.
\beqa
&<&\hat{U}\cdot U_{0,\theta}>\partial_{t'}P_2
-<\hat{U}\cdot DU_{0,\theta}>\nabla^2_{\perp}P_2 \non \\
&-&<\hat{U}\cdot DU_{0,2\theta}>(2\partial_{x'}P_2
+|\nabla_{\perp}P_1|^2)=0,\lb{incomp}
\eeqa
where $\hat{U}$ is an adjoint function of a null eigenfunction of $L$
and $<\hat{U}\cdot U>=\int_{-\infty}^{\infty}(\hat{U}\cdot U)d\theta$.\\
Now, a secular solution up to $O(\eps^2)$ is
\beq
U=U_0(x-vt+\phi)+\eps P_1(\rvp)U_{0,\theta}+\eps^2\{P_2(t',x',\rvp)
U_{0,\theta}+R_2(\rvp)U_{0,2\theta}\}, \lb{intsol}
\eeq
which should be equated with the renormalized $U_0(x-vt+\phit)$,
where $\phit=\phit(t',x',\rvp)$ is a renormalized phase.
Setting
\beq
\phit(t',x',\rvp)=\phi+\delta(t',x',\rvp), \quad |\delta|\ll|\phi|,
\non
\eeq
we expand  $U_0(x-vt+\phit)$ as
\beq
U_0(x-vt+\phit)=U_0(\theta)+\delta U_{0,\theta}+(\delta^2/2) U_{0,2\theta}
+\cdots. \lb{inex}
\eeq
Equating \fr{intsol} and \fr{inex}, we have a renormaization transformation
\beqa
\delta&=&\phit(t',x',\rvp)-\phi=\eps P_1(\rvp)+\eps^2 P_2(t',x',\rvp),
\lb{inrn}\\
\delta^2&=&2\eps^2 R_2(\rvp),\lb{inrn1}
\eeqa
where \fr{inrn1} and \fr{intr} are consistent with \fr{inrn} up to
 $O(\eps^2)$ if $c_4=0$.
 Following the same procedure as the previous examples,
we rewrite \fr{inrn} as
\beq
\phit(t'+\tau,x'+\xi,\rvp+\etav)=\phi(t',x',\rvp)+
\eps P_1(\etav;t',x',\rvp)+\eps^2 P_2(\tau,\xi,\etav;t',x',\rvp),\lb{inrg}
\eeq
from which we obtain
\beqa
\nabla_{\perp}\phit&=&\eps [\partial_{\etav} ( P_1+\eps P_2)]_0,\quad
\nabla^2_{\perp}\phit=\eps^2 [\partial^2_{\etav} P_2]_0,\non\\
\partial_{x'}\phit&=&\eps^2 [\partial_\xi P_2]_0,\quad
\partial_{t'}\phit=\eps^2 [\partial_\tau P_2]_0,\lb{inlie}
\eeqa
where $[P(\tau,\xi,\etav)]_0=P(0,0,0)$.
Subsitution of the relations \fr{inlie} into \fr{incomp}, we
obtain a phase equation for the interface by replacing $t', x'$ by $t,x$.
\beqa
(\partial_{t}&+&2V\partial_{x})\phit=D_{\perp}\nabla^2_{\perp}\phit
+V|\nabla_{\perp}\phit|^2,\lb{phase}\\
D_{\perp}&=&<\hat{U}\cdot DU_{0,\theta}>/<\hat{U}\cdot U_{0,\theta}>,
\non\\
V&=&<\hat{U}\cdot DU_{0,2\theta}>/<\hat{U}\cdot U_{0,\theta}>.\non
\eeqa

If \fr{rdif1} has a periodically oscillating solution $U_0(-\omega t+\phi)
=U_0(\theta)$, the similar procedure as above leads to the isotropic
 Burgers equation for the phase $\phi$.
 In this case, the linealized operator $\tilde{L}$ in
\fr{inu1} is replaced by $L+D\nabla^2, \quad L=\partial_t-F'\cdot$ and
  secular solutions $U_1, U_2$ up to $O(\eps^2)$ are obtained as
 \beqa
U_1&=&P_1(t,\rv)U_0(\theta),\quad U_2=P_2(t,\rv)U_{0,\theta}
+R_2(t,\rv)U_{0,2\theta},\non\\
P_1&=&\nabla P_1\cdot \rv,\non\\
P_2&=&\partial_tP_2 t+[\nabla P_2]_0\cdot \rv+\pv^{(2)}\cdot \rv\rv,
\non\\
R_2&=&P^2_1/2,\non
\eeqa
where $\pv^{(2)}$ is a $2\times 2$ matrix.
The constraint corresponding to \fr{incomp} is given in terms of an adjoint
null function $\hat{U}$ of $L=\partial_t-F'\cdot$ by
\beq
<\hat{U}\cdot U_{0,\theta}>\partial_{t}P_2
-<\hat{U}\cdot DU_{0,\theta}>\nabla^2P_2
-<\hat{U}\cdot DU_{0,2\theta}>|\nabla P_1|^2=0,
\lb{phcomp}
\eeq
where $<\hat{U}\cdot U>=\int_0^{T}(\hat{U}\cdot U)d\theta$ and $T$
is a period of $U_0$ with respect to $\theta$.
A representation of a renormalization group and its generator
have formally the same form as \fr{inrg} and  \fr{inlie}, where
$t',\nabla_{\perp}$ are replaced by $t,\nabla$ respectively.
From \fr{phcomp} and \fr{inlie}, we obtain the three-dimensional Burgers
equation
\beq
\partial_{t}\phit=D_{\perp}\nabla^2\phit
+V|\nabla\phit|^2,\non
\eeq
which is known as the standard phase equation.

\section{Summary}

Through several examples, the perturbative renormalization  group method
 is shown to be understood as the procedure
to obtain an asymptotic expression of a generator of a renormalization
transformation based on the Lie group.
The present approach provides the following simple recipe for obtaining
an asymptotic form of a RG equation from not only autonomous but also
non-autonomous ODE.\\
(1) Get a secular series solution of a perturbed equation by means of naive
perturbation calculations.\\
(2)Find integral constants, which are renormalized to elliminate all the
secular terms in the perturbed solution and give a renormalization
transformation.\\
(3)Rewrite the renormalization transformation by excuting an arbitrary
shift operation on the independent variable: $t\to t+\tau$ and derive
a representation of the Lie group underlying the reormalization
transformation.\\
(4)By differentiating the representation of the Lie group with respect to
arbitrary $\tau$, we obtain an asymptotic expression of the generator,
which yields an asymptotic RG equation.\\
This procedure is valid for general ODE regardless of a translational
symmetry. When the renormalization transformation is known to have a
translational symmetry in advance, the step (3) i.e.
reduction to the Lie group from the
renormalization transformation becomes trivially simple as implied
in \fr{pdtr}. \\
The above recipe for ODE is also applicable to autonomous PDE
if we choose suitable polynomial kernels of the linearized operator.
First, we should take the lowest-order polynomial, of which degree is one,
 as the leading order secular term. As perturbation calculations proceed to
 higher order, polynomial kernels of higher degrees are included in
 the higher-order secular terms order by order.  Through this procedure,
we uniquely determine suitable polynomial kernels among infinite
number of  kernels of the linearized operator and the step (1)
in the recipe is completed.  There are no problems in the other steps.
Thus, the present Lie-group approach is shown to be consistently
applicable to PDE
and some examples are presented. As  more involved examples, we have
succeeded in derivation of higher-order phase equations such as the
non-isotropic Kuramoto-Sivashinsky equation and the  K-dV-Burgers
 equation from the general reaction-diffusion system \fr{rdif1},
 which will be published elsewhere.
\par
\begin{flushleft}
{\Large \bf Ackowledgement}
\end{flushleft}
One of authers (K.N.) wishes to thank Prof.Y.Oono,University of Illinois
,for valuable discussions about the RG method. We also appreciate
Prof. Y.Nambu, Nagoya University, for his introduction to the Einstein's
gravitational theory.\\
\par


\begin{thebibliography}{99}

\bibitem{cgo0}L.Y.Chen,N.Goldenfeld and Y.Oono, Phys.Rev.Lett.73,1311(1994).
\bibitem{cgo1}L.Y.Chen,N.Goldenfeld and Y.Oono, Phys.Rev.E54,376(1996).
\bibitem{yama}Y.Y.Yamaguchi and Y.Nambu, Prog.Theor.Phys.100,199(1998).
\bibitem{sasa}S. Sasa, Physica D108(1997)45(1997).
\bibitem{mano1}K. Matsuba and K. Nozaki, Phys.Rev.E56,R4926(1997).
\bibitem{kuni}T. Kunihiro,Prog.Theor.Phys.94 ,503(1995).
\bibitem{maru}T. Maruo, K. Nozaki and A.Yosimori, Prog.Theor.Phys.101,243
(1999).
\bibitem{mano2}K. Matsuba and K. Nozaki,J.Phys.Soc.Jpn.66,3315(1997).
\bibitem{tani}T.Taniuti,Suppl.Prog.Theor.Phys.55,1(1977).
\bibitem{nay}A.H.Nayfeh,Perturbation Methods(John Willy and Sons, New York,
1973).
\bibitem{shir}D.V.Shirkov,intern.J.Modern Phys.A3,1321(1988).
\bibitem{nambu} Y.Nambu and Y.Y.Yamaguchi, DPNU-99-09 (preprint).
\bibitem{kau}for example, D.J.Kaup and A.C.Newell,Proc.Roy.Soc.London A361
,413(1978).
\bibitem{kar}for example, V.I.Karpman and E.M.Maslov,Sov.Phys.JETP 46,281
(1977).
\bibitem{bon}for example, A.Bondeson,M.Lisak and D.Anderson,Physica Scripta
20,479(1979).
\bibitem{ohta}T.Ohta, ~Mathematical Physics in Interface Dynamics
(Nihon-Hyouronsya,Tokyo,1997) in Japanese.
\end{thebibliography}
\end{document}